\documentclass[sigconf]{acmart}
\usepackage[utf8]{inputenc}
\usepackage{graphicx}
\graphicspath{{figures/}}
\DeclareGraphicsExtensions{.pdf,.jpeg,.png}
\usepackage{amsmath}
\DeclareMathOperator*{\argmax}{argmax}
\usepackage{subfigure}
\usepackage{caption}
\DeclareCaptionType{copyrightbox}
\usepackage{booktabs}
\usepackage{url}
\usepackage{comment}
\usepackage{framed}
\usepackage{xspace}
\usepackage{cleveref}

\usepackage[framemethod=tikz]{mdframed}
\usetikzlibrary{shadows}
\usepackage{graphicx}
\newmdenv[
    tikzsetting= {fill=gray!8},
    skipabove=0.4em,
    skipbelow=0.4em,
    linewidth=1pt,
    innerleftmargin=3pt,
    innerrightmargin=3pt,
    innertopmargin=2pt,
    innerbottommargin=2pt,
    linecolor=gray,
    roundcorner=3pt, 
    shadow=true,
    shadowsize=5pt,
    shadowcolor=gray
]{myshadowbox}

\usepackage{tikz}

\newcommand{\mypara}[1]{\vspace{0.05in}\noindent{\underline{\em {\bf #1}}}}
\newcommand{\impara}[1]{\vspace{0.02in}\noindent{\underline{\em #1}}}

\newcommand{\tableshrink}[1]{\vspace{-0.25in}}
\newcommand{\litcode}[1]{{\small\tt #1}}

\newcommand{\dcom}{{\sc DeepCom}\xspace}
\newcommand{\cnn}{{\sc CodeNN}\xspace}
\newcommand{\fcom}{{\sc FunCom}\xspace}
\newcommand{\dstr}{{\sc DocString}\xspace}

\newcounter{RQCounter}

\newcommand{\RQ}[2]{%
\refstepcounter{RQCounter} \label{#1}
	\vspace{0.1in} \noindent \textbf{RQ\arabic{RQCounter}.~#2 \vspace{0.05in}}
	
}

\newcommand{\RSS}[2]{%
\begin{framed}%
\label{#1}
\filbreak
	\noindent
\textbf{\noindent Result {{\bf #1}}:~~\xspace}{\emph {#2}}%
\end{framed}
}

\setcopyright{rightsretained}

\begin{document}

\copyrightyear{2020}
\acmYear{2020}
%\setcopyright{rightsretained}
\acmConference[ASE '20]{35th IEEE/ACM International Conference on
Automated Software Engineering}{September 21--25, 2020}{Virtual Event,
Australia}
\acmBooktitle{35th IEEE/ACM International Conference on Automated
Software Engineering (ASE '20), September 21--25, 2020, Virtual Event,
Australia}\acmDOI{10.1145/3324884.3416546}
\acmISBN{978-1-4503-6768-4/20/09}

\title{Code to Comment ``Translation'':\\
Data, Metrics, Baselining \& Evaluation}
%Analyzing The Data \& The Metrics}
\author{David Gros*, Hariharan Sezhiyan*, Prem Devanbu, Zhou Yu}
\affiliation{%
  \institution{University of California, Davis}
  %\city{Davis}
  %\state{California}
  %\country{USA}
}

\email{{dgros, hsezhiyan, devanbu, joyu}@ucdavis.edu}

% \date{May 2020}
\begin{abstract}
The relationship of comments to code, and in particular, the task of generating useful comments given the code,
has long been of interest. The earliest approaches have been based
on strong syntactic theories of comment-structures, and relied on textual templates.  
More recently, researchers have applied deep-learning methods to this task---specifically, trainable generative translation models which are known to
%sequence-to-sequence 
work very well for Natural Language translation 
(\emph{e.g.,} from German to English). % Researchers  have applied these models
%reported to give
%promising results for 
%applied to 
%for
%the task of 
%generating comments from code. 
We carefully examine
the underlying assumption here: 
%clearly assumes 
that the task of generating comments sufficiently resembles
the task of translating between natural languages, and
so similar models and evaluation metrics could be used. 
%In this paper, we aim to explore this assumption in more detail. 
We analyze several recent code-comment
datasets for this task: \cnn, \dcom,  \fcom, and \dstr. We compare
them with WMT19, a standard  dataset 
frequently used to train  state-of-the-art
natural language translators. We found some
interesting differences between the code-comment data
and the WMT19
natural language data. 
%emerge this 
%
% very simple IR retrieval methods for this task. 
 %We find evidence suggesting that the IR models can match 
 %the results reported for deep learning models, and analyze in more detail how the %peculiarities of the code-comment data-sets
%affect this result. 
Next, we describe and conduct
some studies to calibrate BLEU (which is commonly used as a measure of
comment quality). 
using ``affinity pairs"
of methods, from different projects, in the same project, in the same class, \emph{etc}; 
%when used to measure the validity \& relevance of comments;
Our study
suggests that the current performance on some datasets might need to be improved substantially. 
We also argue that fairly naive information retrieval (IR) methods
do well enough at this task to be considered a reasonable baseline.
Finally, we make some suggestions on how our findings might be used
in future research in this area. 
%Our work suggests some ways 
%software miners might gather better 
%code-comment training data for deep learners. 
\end{abstract}

\maketitle

\section{Introduction}
Programmers add comments to code to help  comprehension. The value of these comments is well understood and accepted. 
A wide variety of comments exist~\cite{padioleau2009listening} in code, including prefix comments (standardized in frameworks
like Javadocs~\cite{kramer1999api}) which are inserted before functions or methods or modules, to describe their function. Given the value of comments, and the effort required to write them,
there has been considerable interest in providing automated
assistance to help developers to produce comments, and
a variety of approaches have been proposed~\cite{wong2015clocom,mcburney2014automatic,sridhara2010towards,rodeghero2014improving}. 

\footnote{* Authors contributed equally}

\begin{figure}[t]
    \centering
    \includegraphics[width=0.8\columnwidth]{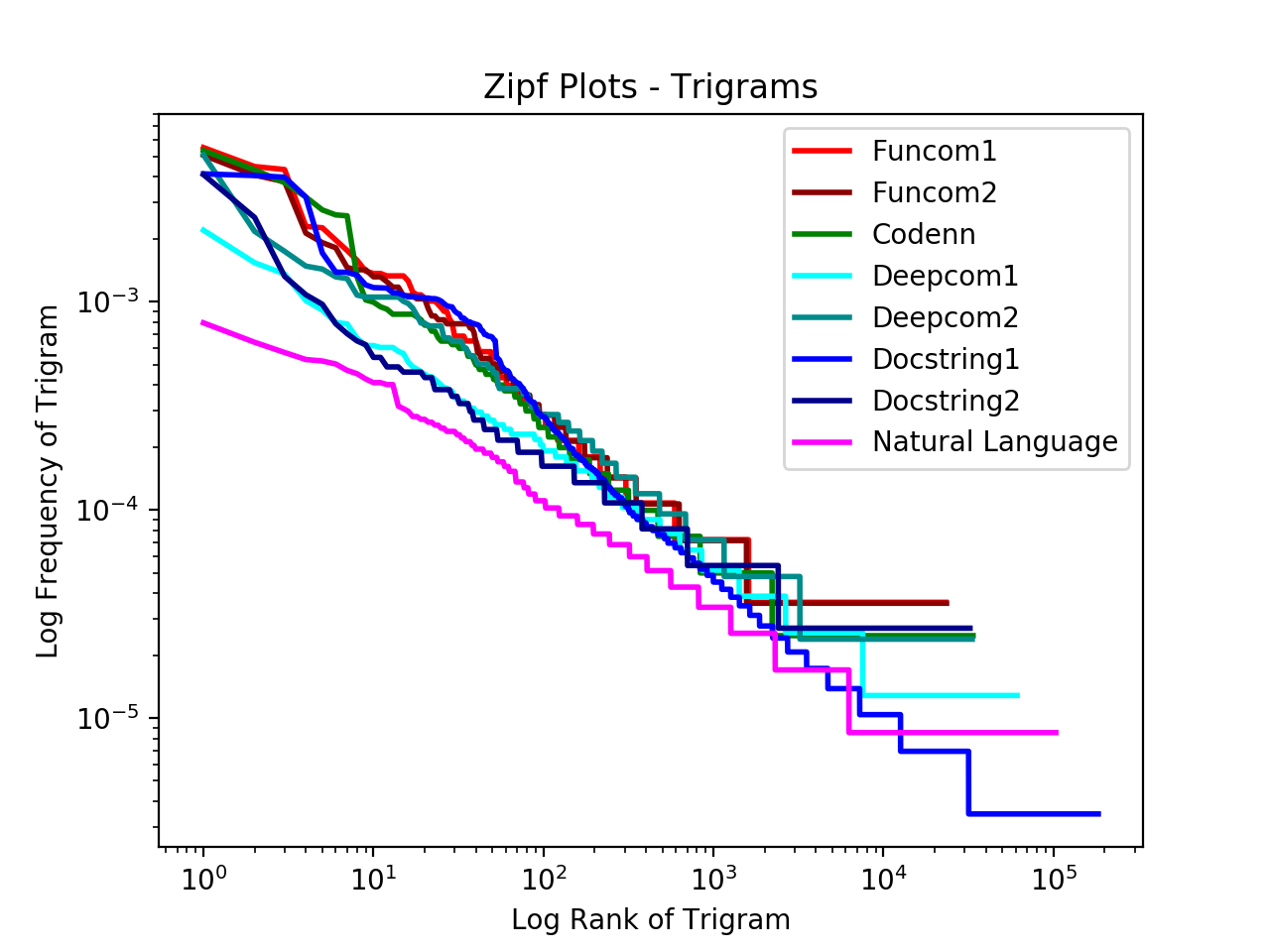}
    \caption{Distribution of trigrams in English (blue) in the WMT~\cite{bojar2014findings}
    German-English machine-translation dataset, and in English  comments from several previously published 
    Code-Comment datasets}
    \label{fig:firstzipf}
\end{figure}
Comments (especially prefix comments) are typically expected to be a useful summary of the function of the accompanying code. Comments
could be viewed as a restatement of the semantics of the code, in a different and more accessible natural language; thus, it is possible to view comment
generation as a kind of translation task, translating from one
(programming) language to a another (natural) language. 
This view, together with the very large volumes of code
(with accompanying comments) available in open-source projects, 
offers the very appealing possibility of leveraging decades
of research in statistical  natural language translation (NLT). If it's possible
to learn to translate from one language to another from data, why not learn
to synthesize comments from code? 
%NLT models are essentially machine learners that learn to translate
%from  language A to language B by leveraging aligned A-B pairs of utterances with the same meaning. 
%More recently, NLT research has been profoundly transformed
%%models~\cite{luong2015effective,klein2017opennmt,vaswani2018tensor2tensor}).
Several recent papers~\cite{hu2018deep,ying2013code, leclair2019neural,iyer2016summarizing} have explored the idea
of applying Statistical Machine Translation (SMT)
methods to \emph{learn} to translate code to an English comments. But are these tasks really similar? We are interested to understand
in more detail how similar the task of generating comments from code is to the task of translating between natural languages. 

%{\underline {\em First,}} while \emph{code} per se is known to be ``natural" in the sense of being repetitive, it is also quite different from natural language~\cite{casalnuovo2019studying}, with simpler grammar, and substantial
%differences in vocabulary sizes and use. Some earlier papers~\cite{hu2017codesum,leclair2019neural} did
%consider these issues. 
Comments form a domain-specific dialect, which is highly structured, with a lot of very
repetitive templates. Comments often begin with patterns like \emph{"returns the"}, \emph{"outputs the"}, and \emph{"calculates the"}. Indeed, most of the earlier work (which wasn't based on 
machine learning) 
on this problem has 
leveraged this highly templated nature of comments~\cite{sridhara2010towards,moreno2013automatic}. 
We can see this phenomenon clearly using Zipf plots. Figure~\ref{fig:firstzipf}  compares the trigram frequencies
of the English language text in comments
(from the datasets~\cite{hu2018deep,leclair2019neural,iyer2016summarizing} that have been used to train deep-learning models for code-comment summarization)
and English language text in the WMT German-English translation 
dataset: the x-axis orders the trigrams from most to least frequent using a log-rank scale,
and the y-axis is the log relative frequency of the trigrams in the corpus. The English
found in WMT dataset is the magenta line at the bottom. The comments
from code show consistently
higher slope in the (note, \emph{log-scaled}) y-axis of the Zipf plot, suggesting
that comments are far more saturated with repeating trigrams than is the English
found in the translation datasets. This observation motivates a closer examination
of the differences between code-comment and WMT datasets, and the implications
of using machine translation approaches for code-comment generation. 

 %{\underline {\em Second,}} in natural language translation, we expect that \emph{more similar inputs will produce more similar outputs}. For the code-commenting task, given that there may be disparate
%ways to implement the same taskthis property may not hold, and may actually affect performance. 

In this paper, we compare code-comment translation (CCT) datasets used with DL models for the task of comment generation, with  a popular natural translation (WMT) dataset used for training DL models for natural language translation. These were our results: 
\begin{enumerate}
    \item We find that the desired outputs for the CCT task are much more repetitive. 
    \item We find that the repetitiveness has a very strong effect on measured performance, much more
    so in the CCT datasets than the WMT dataset. 
    \item We find that the WMT translation dataset has a smoother, more robust input-output dependency. Similar German inputs in WMT have a strong tendency to produce similar English outputs. However, this does appear to hold in the CCT datasets. 
    \item We report that a naive  Information retrieval approach can meet or exceed reported numbers from neural models.
    \item We evaluate BLEU \emph{per se} as a measure of generated comment quality using 
    groups of methods of varying "affinity"; this offers new perspectives on the BLEU  measure. 
\end{enumerate}
%\todo{Revise this bit}
Our findings have several implications for the future work in the area, in terms
of technical approaches, ways of measurement, for baselining, and for calibrating 
BLEU scores. We begin below by first providing some background; we then describe
the datasets used in prior work. We then present an analysis of the datasets and 
and an analysis of the evaluation metrics and baselines used. We conclude
after a detailed discussion of the implications of this work. 

\vspace{0.013in}

\noindent\underline{\emph{But first, a disclaimer:}} this work does not offer any new models for or improvements on prior 
results on the CCT task. 
It is primarily retrospective, \emph{viz}, a critical
review of materials \& evaluations used in prior work in CCT, offered in a collegial spirit, hoping to advance
the way our community views the task of code-comment translation, and
how we might together make further advances in the measurement and evaluation
of innovations that are addressed in this task. 

%Our findings have a few implications. First, they suggest IR approaches might be a useful complement to deep learning approaches in future work\dgr{cite the retrieve papers}. Second, they suggest that duplicates are an important consideration in the development of datasets and evaluating their performance (see also~\cite{allamanis2019adverse}). Third, given the
%differences in data distribution, it may be that alternative architectures for comment generation (like the one used in~\cite{leclair2019neural}) might be very useful. Finally, we note some variations in the
%way BLEU is measured, and suggest a more consistent approach, as noted in~\cite{post2018call}. 

%\dgr{possibly mention reproducability in this field}

\section{Background \& Theory}

The value of comments in code comprehension has been well-established~\cite{takang1996effects}. However, developers
find it challenging to create \& maintain useful comments~\cite{de2005study,Fluri2007DoCA}. This has sparked a long line of research looking into the problem of comment generation. 
An early line of work~\cite{sridhara2011automatically,sridhara2010towards,moreno2013automatic,buse2008automatic}
was rule-based, combining some form analysis of the source code to extract specific information, which could
then be slotted into different types of templates to produce comments. Another approach was
to use code-clone identification to produce comments for given code, using the comments associated
with a clone~\cite{wong2015clocom}.
Other approaches used keywords which programmers seem to 
attend to in eye-tracking studies~\cite{rodeghero2014improving}. Still other approaches use topic analysis to organize descriptions 
of code~\cite{mcburney2014improving}. 

Most of the pioneeering approaches above
relied on specific features and rules hand-engineered for the task of
comment generation. 
The advent of large open-source repositories with large volumes of source-code offered a novel, general, statistically rigorous, 
possibility:  that these large datasets be mined for code-comment pairs, which could then
be used to train a model to produce comments from code. The success of classic statistical
machine translation~\cite{koehn2009statistical} offered a tempting preview of this: using
large amounts of aligned pairs of utterances in languages $A$ \& $B$, it was possible to learn a
conditional distribution of the form $p_t(b \mid a)$, where $a \in A,~$ and $~ b \in B$; given 
an utterance $\beta \in B$, one could produce a possible translation $\alpha \in A$ by simply setting
\[ \alpha = \argmax_a p_t(a \mid \beta) \]
Statistical natural language translation approaches, which were already highly performant, 
were further enhanced by deep-learning (DL). 
Rather
than relying on specific inductive biases like phrase-structures in the case of classical SMT, DL
held the promise that the features relevant to translation
could themselves be learned from large volumes of data. 
DL approaches have led to phenomenal 
improvements in translation quality~\cite{vaswani2018tensor2tensor,klein2017opennmt}. 
Several recent papers ~\cite{iyer2016summarizing,hu2017codesum,leclair2019neural} have explored
using these powerful DL approaches to the code-comment task.

%\pd{Start with templates, lead into DL methods--look at DL papers for motivation, then lead into 
%RQ1: repetitino in output
%RQ2: dependence on input
%RQ3: (report repetition) performance impact of duplication \& repetition
%RQ4: IR results }

%Applying deep learning to generate high level natural language descriptions for code is a popular task in %software engineering.
Iyer \textit{et al.}~\cite{iyer2016summarizing} first applied DL to this task, 
using code-English pairs mined from Stack Overflow---using simple attention over input code, 
and an LSTM to generate outputs. Many other papers followed, which are
discussed below in~\cref{sec:related}. 
%In this empirical study, we show that these code-comment data sets are intrinsically different from natural language data sets in two ways. First, the reference outputs from code-comment data sets are closer in distribution by BLEU than the corresponding reference outputs in natural language data sets. Second, code-comment data sets have a lower linear correlation between the BLEU distribution of inputs and outputs than do natural language NMT data sets. This indicates code comments are based on a fewer set of templates, and in general, are of less variety. To illustrate, we apply information retrieval (IR) on all three data sets, and show the results are comparable with those of deep learning models. The success of IR encourages further discussion about framing comment generation as a translation task and the quality of the mentioned data sets. 
We analyze the published literature, starting with the question of whether there 
are notable distributional differences between the code-comment translation (CCT) and the statistical machine translation (WMT) data. 
Our studies examine the distributions of the input and output data, and the dependence of the output on the input. 

\label{sec:bg}

\RQ{rq:dist}{What are the differences between the translation (WMT) data, and code-comment (CCT) data?}

Next, we examine whether these differences actually affect the performance of translation models. In earlier work, Allamanis~\cite{allamanis2019adverse} pointed out the effects of data duplication on machine learning applications in software engineering. We study the effects of data duplication, as well as the effects of distributional differences on deep learning models. 
One important aspect of SMT datasets is \emph{input-ouput dependence}. In translation \emph{e.g.} from German (DE) to English (EN), similar input DE sentences will to produce similar output EN sentences, and less similar DE sentences will tend
to produce less similar EN sentences. This same correlation might not apply in CCT datasets.

\RQ{rq:perf}{How the distributional differences in the SMT \& CCT data\-sets affect the measured performance?}

%Comments, even for rather different code bodies, often tend to be quite similar; for example 
%The introductory Figure~\ref{fig:firstzipf}, and 
There's another important difference between code and natural language. Small differences, such
as substituting $*$ for $+$ and a  $1$ for a  $0$, can make the difference between a \emph{sum} 
and a \emph{factorial} function; likewise changing one function identifier (\emph{mean}, rather than \emph{variance}). 
These small changes should result in a large change in the associated comment. Likewise, there
are many different ways to write a sort function, all of which might entail the same comment. 
Intuitively, this would appear to be less of an issue in natural languages; since as they have evolved
for consequential 
communication in noisy environments, meaning should be robust to small changes. 
Thus on the whole, we might expect that small changes in German should in general result in only
small changes in the English translation. Code, on the other hand, being a \emph{fiat} language, 
might not be in general as robust, and so small changes in code may result in unpredictable changes in the
associated comment. Why does this matter? In general, modern machine translation methods use the generalized
function-approximation capability of deep-learning models. If natural language translation (WMT) has 
a more functional dependency, and CCT doesn't, there is a suggestion that deep-learning models would find CCT a
greater challenge. 

\RQ{rq:bivar}{Do similar inputs produce similar outputs in both WMT and CCT datasets?}

Prior work in natural language generation has shown that information retrieval (IR) methods can be effective
ways of producing suitable outputs. These methods match a new  input with semantically similar inputs in the training data, and return the associated output. These approaches can sometimes perform quite well~\cite{henderson2019polyresponse} and has been previously applied successfully to the task of comment generation \cite{10.1145/3238147.3240471, Zhang2020RetrievalbasedNS}.
Our goal here is to ask whether IR methods could be a relevant, useful baseline for CCT tasks.

\RQ{rq:IR}{How do the performance of naive Information Retrieval (IR) methods compare across WMT \& CCT datasets? }

Finally, we critically evaluate the use of BLEU scores in this task. Given the differences we found between
datasets used for training SMT translators and the code-comment datasets, we felt it would be important
to understand how BLEU is used in this task, and develop some empirical baselines to calibrate the observed
BLEU values in prior work. How good are the best-in-class BLEU scores (associated with the best current methods 
for generating comments given the source of a method)? Are they only as good as simply retrieving a
comment associated with a random method in a different project? Hopefully they're much better.
How about the comment associated with a random method 
from the same project? With a random method in the same class? 
With a method that could reasonably be assumed quite similar?

\RQ{rq:calib}{How has BLEU been used in prior work for the code-comment task, and how should we view the measured performance?}

In the next section, we review the datasets that we use in our study. 

\section{Datasets Used}

We examine the characteristics of four CCT data sets, namely CodeNN, DeepCom,  FunCom, \& DocString and one standard, widely-used machine-translation dataset, the WMT dataset. 
We begin with a description of each dataset. Within some of the CCT datasets, we observe that the more popular ones can include several different variations: this is because follow-on work has sometimes gathered, processed, and
partitioned (training/validation/test)  the dataset differently. 
%due to wdifferent later authors processing and splitting the datasets differently.
%In particular, we use BLEU as a metric to compare the variation of tokens both in the inputs and outputs of comments and natural language. For comments, we used three datasets: CodeNN, DeepCom, and FunCom. For natural language, we used English and German found in the Workshop on Machine Translation (WMT) 2013 commoncrawl data set.

\label{sec:thedata}
\mypara{CodeNN} Iyer \emph{et al}~\cite{iyer2016summarizing}  was an early CCT dataset, collected from StackOverflow, with code-comment
pairs for C\# and SQL. Stackoverflow posts consist of a title, a question, and a set of
answers which may contain code snippets. 
Each pair consists of the \emph{title} and \emph{code snippet} from answers.
Iyer \emph{et al} gathered around a million pairs each for C\# and SQL; from these, focusing on just snippets in \emph{accepted} answers, they filtered down to 145,841 pairs for C\#  and 41,340 pairs for SQL. From these, they used a trained model (trained using a hand-labeled set) to filter out uninformative titles (\emph{e.g.,} ``How can make this complicated
query simpler") to 66,015 higher-quality pairs for C\# and 33,237 for SQL. In our analysis, we used only the C\# data. 
 StackOverflow  has a well-known community norm to avoid redundant Q\&A; repeated questions are typically referred to the earlier post. As a result, this dataset has \emph{significantly less duplication}. The other  CCT datasets are different. 

\mypara{DeepCom} \citet{hu2018deep} generate a CCT dataset by mining 9,714 Java projects. From this dataset, they filter out methods that have Javadoc comments, and select only those that have at least one-word descriptions. They also exclude getters, setters, constructors and test methods. This leaves them with 69,708 method-comment pairs. In this dataset, the methods (code) 
are represented as serialized ASTs after parsing by Eclipse JDT. 

Later, \citet{Hu2019DeepCC} updated their dataset and model, to a size of 588,108 examples. We refer to the former as DeepCom1 and obtain a copy online from followup work\footnote{\label{neuralcodesum}\url{https://github.com/wasiahmad/NeuralCodeSum/tree/d563e58/data}}. We refer to the latter as DeepCom2 and obtain a copy online\footnote{\url{https://github.com/xing-hu/EMSE-DeepCom/tree/98bd6a}}. In addition DeepCom2 is distributed with a 10-fold split in the cross-project setting (examples in the test set are from different projects). In \citet{Hu2019DeepCC} this is referred to the "RQ-4 split", but to avoid confusion with our research questions, we refer to it as DeepCom2f. 

\mypara{Funcom}
\citet{leclair2019neural} started with the Sourcerer~\cite{bajracharya2006sourcerer} repo, with over 51M methods from 50K projects. 
From this, they filtered out methods with Javadoc comments in English, and then also
the comments that were auto-generated. This leaves about 2.1M methods with patched
Javadoc comments. The source code was parsed into an AST. 
They created two datasets, the \emph{standard}, which retained the original identifiers,
and \emph{challenge}, wherein the identifiers (except for Java API class names) were
replaced with a standardized token. They also made sure no data from
the same project was duplicated across training and/or validation and/or test. Notably, the FunCom dataset only considers the first sentence of the comment. Additionally, code longer than 100 words and comments longer 13 words were truncated. 

Like for DeepCom, there are several versions of this dataset. We consider a version from \citet{leclair2019neural} as FunCom1 \ and the version from \citet{DBLP:journals/corr/abs-1904-02660} as FunCom2. These datasets are nearly identical, but FunCom2 has about 800 fewer examples and the two versions have reshuffled train/test/val splits. The Funcom1\footnote{\url{http://leclair.tech/data/funcom/}} and Funcom2\footnote{\url{https://s3.us-east-2.amazonaws.com/icse2018/index.html}}
datasets are available online. 

\mypara{Docstring} \citet{baroneDocstring} collect Python methods and prefix comment "docstrings" by scraping GitHub. Tokenization was done using subword tokenization. They filtered the data for duplications, and also removed excessively long examples (greater than 400 tokens).  However, unlike other datasets, Barone \emph{et al.} do not limit to only the first sentence of the comments. This can result in relatively long desired outputs.

The dataset contains approximately 100k examples, but after filtering out very long samples,
as per Barone \emph{et al} preprocessing script\footnote{\url{https://bit.ly/2yDnHcS}}, this is reduced to 74,860 examples. We refer to this version as DocString1.

We also consider a processed version obtained from \citet{ahmad2020transformerbased} source\textsuperscript{\ref{neuralcodesum}} which was attributed to \citet{wei2019code}. We refer to this version as DocString2. Due to the processing choices, the examples in DocString2 are significantly shorter than DocString1.

\mypara{WMT19 News Dataset}
To benchmark the comment data with natural language, we used data from the Fourth Conference of Machine Translation (WMT19). In particular, we used the news dataset~\cite{barrault-etal-2019-findings}. After manual inspection, we determined this dataset offers a good balance of formal language that is somewhat domain specific to more loose language common in everyday speech. In benchmarking comment data with natural language, we wanted to ensure variety in the words and expressions used to avoid biasing results. We used the English-German translation dataset, and compared English in this dataset to comments in the other datasets (which were all in English) to ensure differences in metrics were not a result of differences in language.

%\mypara{Standardization} To ensure reproducibility of our experiments, we adopted a few guidelines. 
%First, we tokenize the comments data sets using the same tokenization techniques used by the original designers of those data sets. For example, the designers of Funcom split solely by space characters, while the designers of CodeNN apply a more involved tokenization which separates punctuation and special characters. Second, we ensure the data sets have a similar vocabulary to accurately depict the data. If one data set is considerably larger, its intrinsically larger vocabulary would artificially deflate the frequency of the most common ngrams and skew the plot downwards. Finally, in the case of Funcom, we removed sentence start (<s>) and close (</s>) tokens.

\impara{Other CCT Datasets}
We tried to  
capture most of the code-comment datasets that are used in the context of translation. However, there are some recent datasets which could be used in this context, but we did not explore \cite{husain2019codesearchnet, agashe2019juice}. While doing our work we noticed that some prior works provide the raw collection of code-comments for download, but not the exact processing and evaluations used~\cite{mooreConvComment}. Other works use published datasets like DocString, but processing and evaluation techniques are not now readily available~\cite{wang2020trans3, 9031440}. As we will discuss, unless the precise processing and evaluation code is available, the results may be difficult to compare. 

% Essentially need to go through and explain each of the 
% Define 

\subsection{Evaluation Scores Used}
\label{sec:evalused}

A common metric used in evaluating text generation is BLEU score \citep{papineni2002bleu}. When comparing translations of natural language, BLEU score has been shown to correlate well with human judgements of translation quality~\cite{coughlin2003correlating}. In all the datasets we analyzed, the associated papers used
BLEU to evaluate the quality of the comment generation. However, 
%when analyzing the comment generation datasets we find a some issues, namely that while all datasets use a BLEU score, 
there are rather subtle differences in the way the BLEUs were calculated, which makes 
the results rather difficult to compare. We begin this discussion with a
% \subsubsection{BLUE Background}
%Before expanding the differences and distinctions of each dataset's BLEU metric, we first provide a 
brief explanation of the BLEU score.

BLEU (as do related measures) indicates the closeness of a candidate translation output 
to a ``golden" reference result. 
BLEU \emph{per se} measures the \emph{precision} (as opposed to \emph{recall}) of a candidate, relative to the reference,
using constituent
\textit{n}-grams.
%1-grams (unigrams) are single tokens, while 4-grams are of sequences of 4 consecutive tokens. 
BLEU typically uses unigrams through 4-grams to measure the precision of the system output. If we define \textit:
\[ p_n = \frac{\text{number of \textit{n}-grams in both reference and candidate}}{\text{number of \textit{n}-grams in the candidate}} \]

BLEU combines the precision of each \textit{n}-gram using the geometric mean, $\exp({\frac{1}{N}\sum_{n=1}^{N} \log{p_n}) }$. With just this formulation, single word outputs or outputs that repeat common \textit{n}-grams could potentially have high precision. Thus, a \emph{``brevity penalty''} is used to scale the final score; furthermore each \textit{n}-gram in the reference can be used in the calculation just once. \citep{eisenstein2018natural}
These calculations are generally standard in all BLEU implementations, but several variations may arise. 

\impara{Smoothing:} One variation arises when deciding how to deal with cases when $p_n = 0$, \emph{i.e.},  an $n$-gram in the candidate string is not in the reference string \cite{Chen2014ASC}. 
With no adjustment, one has an undefined $\log{0}$. One can add a small epsilon to $p_n$ which removes undefined expressions.
However, because BLEU is a geometric mean of $p_n, n \in \{1, 2, 3, 4\}$ if $p_4$ is only epsilon above zero, it will result in a mean which is near zero. Thus, some implementations opt to smooth the $p_n$ in varying ways. To compare competing tools
for the same task, it would be preferable to use a standard measure. 

\impara{Corpus vs. Sentence BLEU:} 
When evaluating a translation system, one typically measures BLEU 
(candidate \emph{vs} reference) across all the samples in the held-out test set. Thus another source of implementation variation is when deciding how to combine the results between all of the test set scores. 
One option, which was proposed originally in Papineni \textit{et al.}~\cite{papineni2002bleu}, is a "corpus BLEU", sometimes referred to as C-BLEU. In this case the numerator and denominator of $p_n$ are accumulated across every example in the test corpus. This means as long as at least one example has a 4-gram overlap, $p_4$ will not be zero, and thus the geometric mean will not be zero
An alternative option for combining across the test corpus is referred to as "Sentence BLEU" or S-BLEU. In this setting BLEU score for the test set is calculated by simply taking the arithmetic mean the BLEU score calculated on each sentence in the set. 

\impara{Tokenization Choices:} A final source of variation comes not from how the metric is calculated, but from the inputs it is given. Because the precision counts are at a token level, it has been noted that BLEU is highly sensitive to tokenization \citep{post2018clarity}. This means that when comparing to prior work on a dataset, one must be careful not only to use the same BLEU calculation, but also the same tokenization and filtering. When calculating scores on the datasets, we use the tokenization provided with the dataset.

Tokenization can be very significant for the resulting score. As a toy example, suppose a reference contained the string ``{\small\tt calls function foo()}'' and an hypothesis contained the string ``{\small\tt uses function foo()}''. If one chooses to tokenize by spaces, one has tokens [{\small\tt calls}, {\small\tt function}, {\small\tt foo()}] and [{\small\tt uses}, {\small\tt function}, {\small\tt foo()}]. This tokenization yields only one bigram overlap and no trigram or 4-gram overlaps. However, if one instead chooses to tokenize this as [{\small\tt calls}, {\small\tt function}, {\small\tt foo}, {\small\tt (}, {\small\tt )}] and [{\small\tt uses}, {\small\tt function}, {\small\tt foo}, {\small\tt (}, {\small\tt )}] we suddenly have three overlapping bigrams, two overlapping trigrams, and one overlapping 4-gram. This results in a swing of more than 15 BLEU-M2 points or nearly 40 BLEU-DC points (BLEU-M2 and BLEU-DC described below).

%\subsubsection{BLEU Variants}\label{bleu-versions}

We now go through BLEU variants used by each of the datasets and assign a name to them. The name is not intended to be prescriptive or standard, but instead just for later reference in this document. All scores are the "aggregate" measures, which consider up to \textit{4}-grams.

\mypara{BLEU-CN} This is a Sentence BLEU metric. It applies a Laplace-like smoothing by adding 1 to both the numerator and denominator of $p_n$ for $n \geq 2$. The CodeNN authors' implementation was used \footnote{\url{https://github.com/sriniiyer/codenn/blob/0f7fbb8b298a8/src/utils/bleu.py}}.

\mypara{BLEU-DC} This is also a Sentence BLEU metric. The authors' implementation is based off NLTK~\cite{LoperBird02} using its "method 4" smoothing. This smoothing is more complex. It only applies when $p_n$ is zero, and sets  $p_n = 1 / ((n-1) + 5/\log{l_h})$ where $l_h$ is the length of the hypothesis. See the authors' implementation for complete details\footnote{\url{https://github.com/xing-hu/EMSE-DeepCom/blob/98bd6aac/scripts/evaluation.py}}.

\mypara{BLEU-FC} This is an unsmoothed corpus BLEU metric based on NLTK's implementation. Details are omitted for brevity, and can be found in the authors' source\footnote{\url{https://github.com/mcmillco/funcom/blob/41c737903/bleu.py\#L17}}.

\mypara{BLEU-Moses} The Docstring dataset uses a BLEU implementation by the Moses project\footnote{\url{https://bit.ly/2YF0hye}}. It is also an unsmoothed corpus BLEU. This is very similar to BLEU-FC (though note that due to differences in tokenization, scores presented by the two datasets are not directly comparable).

\mypara{BLEU-ncs} This is a sentence BLEU used in the implementation\footnote{\url{https://github.com/wasiahmad/NeuralCodeSum/blob/b2652e2/main/test.py\#L324}} of Ahmad et al. \cite{ahmad2020transformerbased}. Like BLEU-CN, it uses an add-one Laplace smoothing. However, it is subtly different than BLEU-CN as the add-one applies even for unigrams.

\mypara{SacreBLEU} The SacreBLEU implementation was created by \citet{post2018clarity} in an effort to help provide a standard BLEU implementation for evaluating on NL translation. We use the default settings which is a corpus BLEU metric with an exponential smoothing.

\mypara{BLEU-M2} This is a Sentence BLEU metric based on nltk "method 2" smoothing. Like BLEU-CN it uses a laplace-like add-one smoothing. This BLEU is later presented in plots for this paper.

\smallskip
We conclude by noting that the wide variety of BLEU measures used in prior work in code-comment translation carry some risks. We discuss further
below. \cref{tab:diffBLEU} provide some evidence suggesting that the variation is high enough to raise some concern about the true interpretation of claimed advances; as we argue below, the field can benefit from further standardization. 

\subsection{Models \& Techniques}
~\label{sec:related}
In this section, we outline the various deep learning approaches that have been 
applied to this code-comment task. We note that our goal in this paper is not to critique or improve upon
the specific technical methods, but to analyze the data \emph{per se} to gain some insights on
the distributions therein, and also to understand the most comment metric (BLEU) that is used, 
and the implications of using this metric. However, for completeness, we list
the different approaches, and provide just a very brief overview of each technical
 approach. 
All the datasets used below are described above in~\cref{sec:thedata}. 

Iyer \emph{et al}~\cite{iyer2016summarizing} was an early attempt at this task, using a fairly standard seq2seq RNN model, enhanced with attention. Hu \emph{et al}~\cite{hu2018deep} also
used a similar RNN-based seq2seq model, but introduced a  ``tree-like" preprocessing of the input
source code. Rather than simply streaming in the raw tokens, they first parse it, and then serialize the resulting AST into a token stream
that is fed into the seq2seq model. A related approach~\cite{alon2018code2seq} digests a fixed-size random sample of paths through the AST of the input code (using LSTMs) and produces code
summaries. LeClair \emph{et al}~\cite{leclair2019neural} proposed an approach that combines
both structural and sequential representations of code; they have
also suggested the use of graph neural networks~\cite{alex2020improved}.  Wan \emph{et al}~\cite{wan2018improving} use a similar approach, but advocate using reinforcement learning to enhance the generation element. More recently, the use of function context~\cite{haque2020improved} has been reported to improve comment synthesis. Source-code vocabulary 
proliferation is a well-known problem~\cite{karampatsis2020big}; previously unseen identifier or method names
in input code or output comments can diminish performance. New work by Moore \emph{et al}~\cite{mooreConvComment} approaches
this problem by using convolutions over individual \emph{letters} in the input and using 
subtokens (by camel-case splitting) on the output. Very recently~\citet{Zhang2020RetrievalbasedNS} have
reported that combining sophisticated IR methods with deep-learning leads to further gains in the CCT task. For our purposes (showing that IR methods constitute a reasonable baseline) we use a very simple, vanilla, out-of-box Lucene
IR implementation, which already achieves nearly SOTA performance in many cases. 

There are tasks related to generating comments from code: for example, synthesizing a commit
log given a code change~\cite{cortes2014automatically,liu2018neural,jiang2017automatically}, or generating
method names from the code~\cite{allamanis2016convolutional,alon2018code2seq}. Since these
are somewhat different tasks, with different data characteristics, we don't discuss them further. 
In addition 
code synthesis~\cite{yin2018learning,agashe2019juice} also uses matched pairs of natural language
and code; however, these datasets have not been used for generating English from code, and
are not used in prior work for this task; so we don't discuss them further here. 
%Tree to sequence: 
%Training methods: teacher forcing+ autoregressive vs. RL based training

%\todo{talk also about MUSE. }\cite{mooreConvComment}

\section{Methods \&  Findings}
In the following section, we present our methods and results for each of the  RQs presented in \S~\ref{sec:bg}. In each case, we present some illustrative plots and (when applicable) the results of relevant statistical tests. All p-values have been corrected using family-wise (Benjamini-Hochberg) correction. 
To examine the characteristics of each dataset, we constructed two types of plots: zipf plots and bivariate BLEU plots. 
\subsection{Differences between CCT and WMT data}
%\prem{Outline Zipf plots, show a couple }

\begin{figure}[h]
%\centering
   %   \subfigure[]{\includegraphics[width=0.48\columnwidth]{other_plots/unigrams_smoothed.png}}
%\subfigure[]{\includegraphics[width=0.48\columnwidth]{other_plots/trigrams_smoothed.png}}
\includegraphics[width=0.74\columnwidth]{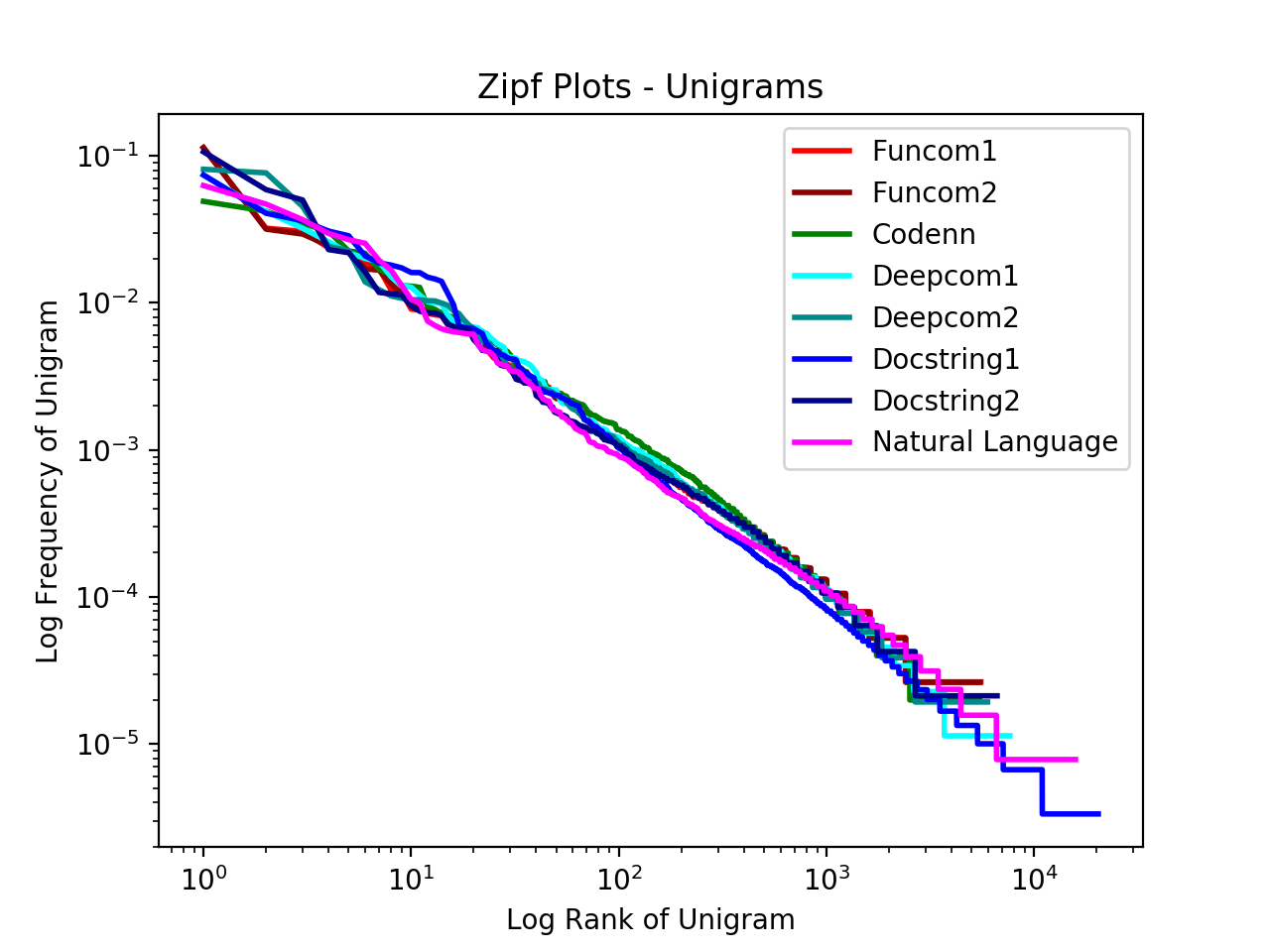}
\caption{Unigram (Vocabulary) distribution. Zipf plot for all datasets look similar. Difference from trigram Zipf
plot in Fig~\ref{fig:firstzipf} suggests greater repetitiveness in code comments.}
\label{fig:zipf1}       
\end{figure}
The Zipf plots are a useful way to visualize
the skewness of textual data, where (in natural text) a few tokens (or ngrams) account for a large portion of the text. Each plot point is a (rank, relative-frequency) pair, both log-scaled. We use the plot to compare the relative skewness of the (English) comment data in the CCT data and the desired English outputs in the WMT NLT data. Examining the unigram
 Zipf plot above, it can be seen in both code comments and natural English, a few vocabulary words do  dominate. However, when we turn back to the trigram Zipf plots in Figure~\ref{fig:firstzipf}, we can see the
%As can be seen in the introductory Figure~\ref{fig:firstzipf} and in 
difference.
%This indicates that the most frequent trigrams in code comments are substantially more frequent; 
%this  indicates that code comments are much more stylized and templated than
%natural English (blue line at bottom, with shallowest slope)
%with some sequences of tokens occurring much more often. 
%Figure~\ref{fig:zipf1} shows the \emph{unigram} (vocabulary) distributions
%on the English comments across 4 CCT and the English part of WMT German-English dataset.
%Comparing this plot with the trigram distributions 
%in Figure~\ref{fig:firstzipf}, 
One is left with the clear
suggestion that: while the vocabulary distributions across the different datasets 
aren't that different, the ways in which these vocabulary words
are \emph{combined into trigrams} are much more stylistic and templated in code comments. 

\RSS{1}{Code comments are far more repetitive than the English found in Natural Language Translation datasets}

Given this relatively greater repetitive structure in code comments, we can expect that the performance
of translation tools will be strongly influenced by repeating (and/or very frequent) trigrams. If a few
frequent $n$-grams account for most of the desired output in a corpus, it would seem that these trigrams
would play a substantial, perhaps misleading role in measured performance.
\begin{figure*}[]
%\centering
   %   \subfigure[]{\includegrarm -phics[width=0.48\columnwidth]{other_plots/unigrams_smoothed.png}}
%\subfigure[]{\includegraphics[width=0.48\columnwidth]{other_plots/trigrams_smoothed.png}}
\includegraphics[width=1.25\columnwidth]{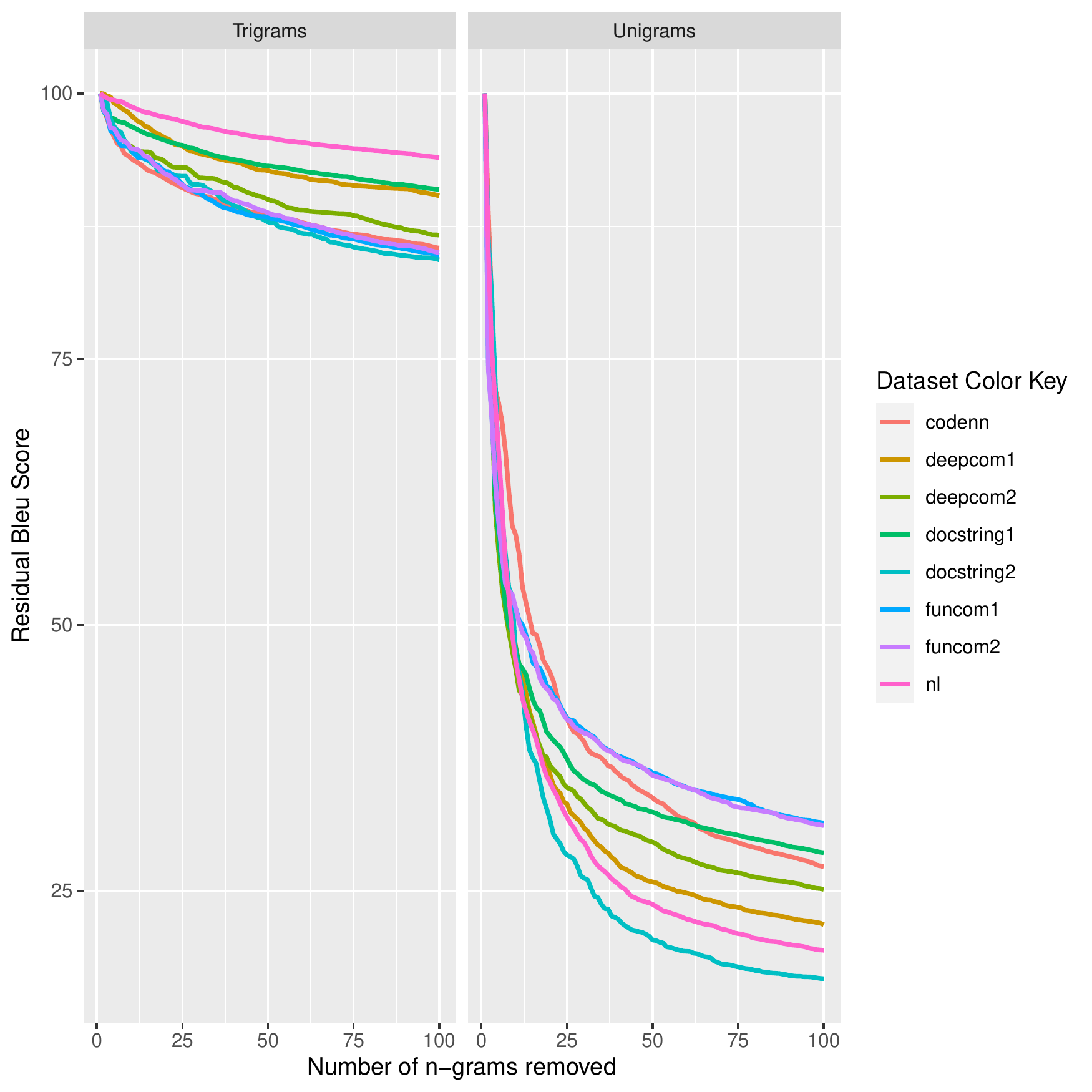}
% Original
%\includegraphics[width=0.98\columnwidth]{other_plots/dropoff.png}
\caption{The effect of most frequent unigrams (left) and trigrams (right) on measured (smoothed) BLEU-4 performance. BLEU-4 was calculated after successive removals of of most frequent unigrams (right) and trigrams (left). The effect of removing frequent \emph{Unigrams} is by and large greater on the natural language dataset ("nl"). 
However, the effect of removing frequent \emph{trigrams}, on the comment datasets is generally stronger than on the "nl" dataset  due to high degree of repetition in the comment datasets. These apparent
visual differences are decisively confirmed by more rigorous statistical modeling.}
%%%%

\label{fig:removal}       
\end{figure*}
%\hsz{Hari, we could make the text in these figures bigger without affecting the fit, I think. Let's do that}
%\pd{Add a description of the changing scales between the unigram and trigram}
Figure~\ref{fig:removal} supports this analysis. The right-hand side plot shows the effect on BLEU-4 
of replacing single words (unigrams) with
random tokens in the ``Golden" (desired) output in the various datasets. The left-hand plot shows
the effect of replacing trigrams. The index (1 to 100) on the x-axis shows the number
of most frequent n-grams replaced with random tokens. The y-axis shows the decrease in measured
BLEU-4 as the code is increasingly randomized. 

The Unigrams plot suggests that the effect on the desired Natural language  ("nl") output,
as measured by BLEU is relatively greater when compared to most of the comment datasets. 
This effect is reversed for Trigrams; the "nl" dataset is not affected as much by
the removal of frequent Trigrams as the comment datasets.
This analysis suggests that a tool that got the top few most frequent
trigrams wrong in the code-comment generation task would suffer a larger performance penalty than  a tool that
got the top-few $n$-grams wrong in a natural language  translation task. This visual evidence is strongly confirmed by rigorous statistical modeling, please see supplementary materials, bleu-cc.Rmd for the R code. 
Frequent trigrams that have a big effect on the code comment BLEU include \emph{e.g.,} ``{\small\tt factory method for}'', 
``{\small\tt delegates to the}'', and ``{\small\tt method for instantiating}''. 
To put it another
way, one could boost the performance of a code-comment translation tool, perhaps misleadingly, by getting
a few such $n$-grams right. 

\begin{comment}
To get a quantitative, statistical sense of the greater degree of repetition in the desired outputs in CCT datasets relative to NLT datasets, we measure the similarity between desired outputs using BLEU-1. BLEU-1 is a normalized measure of the number of overlapping unigrams (tokens), with 0 representing no similarity and 1 representing perfect overlap in tokens. BLEU-2 is like BLEU-1, but for bigrams, and so on.  
To get a fair sample of similarities, we randomly select 100,000 pairs of outputs, from each dataset. For each pair, we compute the similarity using BLEU-1. Since the datasets aim to be free of duplicates, 
generally we find a fair number of neglibible BLEU-1 values; for BLEU-2 and beyond most pairs show no similarity. 
\end{comment}
%\pd{add plots and discuss}.

\RSS{2}{Frequent n-grams could wield a much stronger effect on the measured BLEU performance on code-comment translation tasks than on natural language translation}

\subsection{Input-Output Similarity (RQ3)}

\begin{figure*}[]
  \includegraphics[width=2.05\columnwidth]{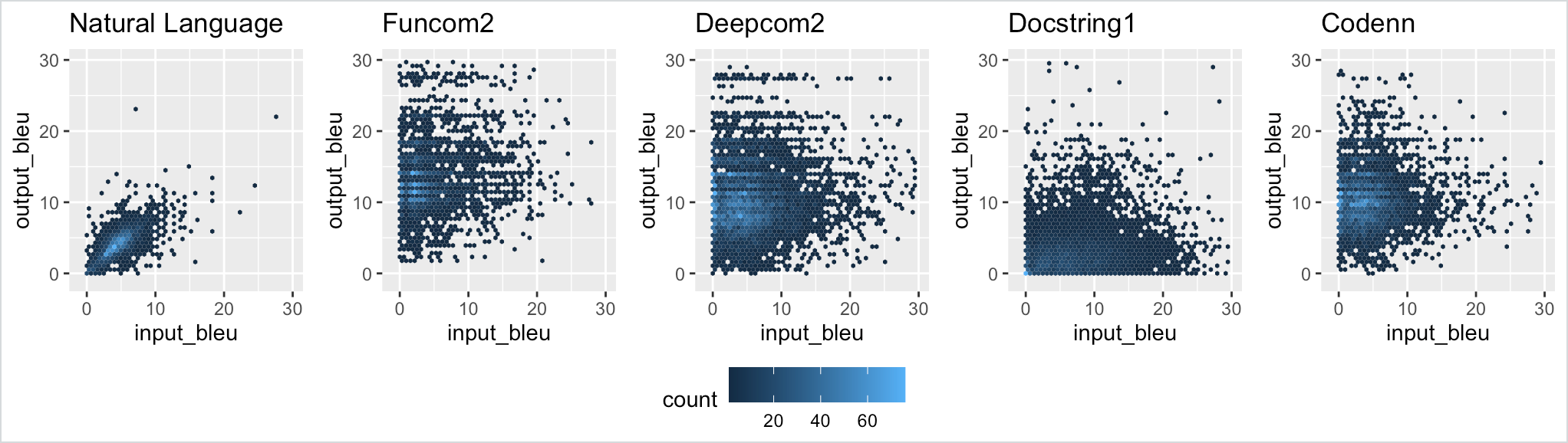}
     %\caption{Zipf plot of \emph{unigram} distributions of the same datasets as in Figure~\ref{fig:firstzipf}}

         \caption{Bivariate plots showing dependency of input-similarity to output-similarity. Pairs with similarity (BLEU-4) less than $10^{-5}$ are omitted. The natural language translation data have the strongest dependency: similar inputs have the strongest tendency to provide similar outputs}
\label{fig:bivariate}         
\end{figure*}

An important property of natural language translation is that there is a general \emph{dependence of
input on output}. Thus, similar German sentences should translate to similar English sentences. For example
two German sentences with similar grammatical structure and vocabulary should in general result in two English
sentences whose grammatical structure and vocabulary resemble each other; likewise, in general, the more different two German sentences are in vocabulary and grammar, the more difference we expect in their English translations. Exceptions
are possible, since some similar constructions have different meanings: 
(\emph{kicking the ball} vs. \emph{kicking the bucket}\footnote{The latter idiom indicates death;  Some automated translation engines (\emph{e.g.} Bing) seem to know the difference when translating from English}
). However, on average in large datasets, we should expect that more similar sentences give more similar translations. 

When training a translation engine with a high-dimensional non-linear
function approximator like an encoder-decoder model using deep learning, this monotonic
dependence property is arguably useful. We would expect similar input sentences to be encode into similar
points in vector space, thus yielding more similar output sentences. How do natural language
translation (German-English) and code-comment datasets fare in this regard? To gauge this 
phenomenon, we sampled 10,000 \emph{random pairs of input fragments} from each of 
our datasets, and measured 
their similarity using BLEU-M2, as well as the similarity of the corresponding Golden (desired)
output fragments. We then plot the \emph{input} BLEU-M2 similarity for each sampled pair
on the x-axis, and the  
BLEU-M2 similarity of the corresponding pair of \emph{outputs} on the y-axis. We use a kernel-smoothed
2-d histogram %(a smoothed Hexbin plot)
rather than a scatter plot, to make the frequencies more
visible, along with (we expect) an indication suggesting that similar inputs yield similar outputs. 
Certainly, most inputs and outputs are different, so we expect to find a large number of highly dissimilar 
pairs where 
input and output pair BLEUs are virtually zero. So we considered our random samples, with and
without input and output similarities  bigger than
$\epsilon = 10^{-5}$. The highly dissimilar input-output pairs 
are omitted in the plot. However, as an additional
quantitative test of correlation, we also considered Spearman's $\rho$. 
both with and without the dissimilar pairs. 

\begin{table}[h]
    \centering
    \begin{tabular}{lllc}
        \toprule
        %{} & {} & \multicolumn{2}{c}{BLEU} \\
        %\cmidrule{3-5} \\
        Dataset & \emph{Spearman's} $\rho$ & Significance & \emph{Number of} \\
    
          &\emph{Correlation} &\emph{p-value} & \emph{pairs with}\\
                & $\rho$, $BLEU> \epsilon$  & p-value, $BLEU> \epsilon$  & BLEU $> \epsilon$ \\
                                & ($\rho$,all)  & (p-value, all)  &  \\
        \toprule
        NL & 0.70 (0.02) & 0.0 (0.055) & 2568 \\
        %Deepcom (no 0's) & 0.107 & 1.097e-13 & 4835 \\
        \midrule
        Deepcom1 & 0.056 (0.057) & 1.0e-6 (2.2e-8) & 7811 \\
        %Deepcom (no 0's) & 0.107 & 1.097e-13 & 4835 \\
        \midrule
        Deepcom2 & 0.036 (0.045) & 1.5e-3 (1.1e-5) & 7966 \\
        %Deepcom (no 0's) & 0.107 & 1.097e-13 & 4835 \\
        %Docstring (no 0's) & 0.152 & 2.864e-46 & 5211 \\
        \midrule
        Docstring1 & 0.147 (0.16) & 6.7e-42 (4.6e-59) & 8585 \\
         \midrule
        Docstring2 & 0.041 (0.047) & 9.5e-5 (3.7e-6) & 9585 \\
        %Docstring (no 0's) & 0.152 & 2.864e-46 & 5211 \\
        \midrule
        Funcom1 & 0.124 (0.083) & 1.30e-18 (3.4e-16) & 5026 \\
        % Funcom (no 0's) & 0.0974 & <  0. & 5211 \\
        \midrule
        Funcom2 & 0.122 (0.079) & 3.03e-18 (6.7e-15) & 5082 \\
        % Funcom (no 0's) & 0.0974 & <  0. & 5211 \\
        \midrule
        Codenn & 0.012 (0.0012) & 0.409 (0.904) & 2532 \\
        %Docstring (no 0's) & 0.152 & 2.864e-46 & 5211 \\
        \bottomrule
    \end{tabular}
    \caption{Correlation values (Spearman's $\rho$, and significance, $p$-value,  for the plots  in Figure~\ref{fig:bivariate}. Values outside paranthesis are calculated with only the pairs having  pairwise BLEU $> 10^{-5}$; values in paranthesis include all pairs. 
    $p$-values are adjusted with Benjamini-Hochberg familywise correction. In all cases, we chose 10,000 random
    pairs} 
    \label{tab:all-IR}
\vspace{-0.35in}                    
\end{table}

%\todo{bit confused about what non-zero points are}
The bivariate plots are shown in Fig~\ref{fig:bivariate} and the Spearmans $\rho$ in~\cref{tab:ir-results}.
%The data are based on $n=$10,000 random pairs from each dataset, where we measure
%the BLEU-4 similarity between the inputs and the outputs.
We have split the 10,000 random samples
into two cases: one in which we do no further processing and one where both input/output BLEU similarities are both $> 10^{-5}$. 
%For our purposes, we consider $<= 10^{-5}$ similarity to be zero similarity.
Spearman's $\rho$, 
significance, and sample sizes are shown for non-zeroes in the columns (the numbers within
the parentheses include the highly dissimilar ones). 
From the table (last column) we see that about 25-96\% of the pairs have some similarity on both inputs
and outputs, depending on the dataset.
The table also shows the Spearman's $\rho$ (first column) and significance (second column). 
Each subplot in Fig~\ref{fig:bivariate}  shows one dataset, where
x-axis is the BLEU similarity of a pair's inputs, and y-axis is that of outputs. The plot is a binned 2D histogram, 
using colored hexagons to represent counts in that bin. A representative variant of each dataset is plotted (as applicable); the omitted
ones are visually very similar. 

%First, it is clear the natural language pairs are more closely clustered near the origin. This indicates random pairs in natural language are less likely to have common tokens and ngrams. 
We can  clearly see a stronger relationship between input and output BLEUs in the natural language setting. Particularly
for natural language data, this is further evidenced by the rather high Spearman correlation for
the non-zero BLEU pairs (\emph{0.70!!}), and the evident
visual dependence
between input-input similarity and output-output similarity is note worthy; this indicates
that there is strong, fairly monotonic relationship in natural language translation: the more similar the source,
the more similar the translation! 

This analysis suggests that natural language data has a stronger, more specific input-output dependence; this
also suggests that translation between languages
is more amenable to learnable function-approximators like deep learners; this appears
to be \emph{substantially less true} for code-comment data. 
%, which although does not directly test for linearity, strongly implies it. 
%The linearity implies code-comment datasets have a less clear one-to-one mapping: specific tokens or ngrams in the input do not necessarily correspond to specific tokens or ngrams in the output. This would allow models to predict certain output tokens that naturally achieve higher BLEU scores regardless on input tokens.
This gives us the following conclusion with reference to RQ3. 

\RSS{3}{
The natural language translation (WMT) shows a stronger input-output dependence than the CCT datasets in that
similar inputs are more likely to produce similar outputs.
}

%\hsz{Somethign about how impact on performance was evaluated}

%\dgr{How the IR based "generation" was implemented}

\subsection{Information Retrieval Baselines}
As can be see in ~\cref{fig:bivariate} and ~\cref{tab:ir-results}, datasets for the natural
language translation task show a smoother and more monotonic  input-output dependence;
by contrast, code-comment  datasets seem to have little or no input-output dependence. 
This finding casts some doubt on the existence of a general sequence-to-sequence 
$code \rightarrow comment$ function that
can be learned using a universal function approximator like a deep neural network. However it leaves
open the possibility that a more data-driven approach, that simply memorizes the training
data in some fashion, rather than trying to generalize from it, might also work. Thus, given
a code input, perhaps we can just try to find similar code in the training dataset, and
retrieve the comment associated with the similar code. This is a simple and naive information-retrieval
(IR) approach. We then compare this to the IR performance on NL translation.
%We found IR to be unexpectedly performant; as we detail below,
%we believe IR works well enough to be considered a universal straw-man baseline for this task. 

\subsubsection{Method}
We use Apache Solr Version 8.3.1\footnote{https://lucene.apache.org/solr/}
to implement a straightforward IR approach. Apache Solr is a open source document search engine based on Apache Lucene. We simply construct an index of over the code parts
of the relevant datasets; given a code input, we use that as a ``query" over the index, find
the closest match, and return the comment associated with the closest matching code as
the ``generated comment". 

We used the default parameters of Solr without tuning. This includes the default BM25 scoring function \cite{robertson1994some}. 
For each dataset, we use always the same tokenization procedure used by authors. In addition, 
%Other than the dataset's given tokenization, 
we perform some additional pre-processing on the code, that is typically required
for IR approaches. For example, we remove highly frequent stop words from the code. Additionally, for datasets do not provide a tokenization phase
that actually splits  cammelCaseWords or snake\_case\_words, we include terms for indexing and searching which includes the split form of these words.
However, we note that the processing of stop words and word splitting only effects a minor change in performance.

\begin{table}[h]
    \centering
    \begin{tabular}{lllll}
        \toprule
        %{} & {} & \multicolumn{2}{c}{BLEU} \\
        %\cmidrule{3-5} \\
        Dataset & Method Used & Score & Score Method \\
        \toprule
        
        DeepCom2f & IR-Baseline & 32.7 & BLEU-DC \\
        %%%%%% OverEdit
        %- & DeepCom (SBT) \cite{hu2018deep} & 38.2 \cite{Hu2019DeepCC} & - \\
        %- & Seq2Seq \cite{seq2seq} & 34.9 \cite{Hu2019DeepCC} & - \
        %%%%%%
        - & DeepCom (SBT) \cite{hu2018deep} & 38.2 & - \\
        - & Seq2Seq \cite{seq2seq} & 34.9 \cite{hu2018deep} & - \\
        
        \midrule

        DeepCom1 & IR-Baseline  & {\bf 45.6} & BLEU-ncs \\
        - & Transformer \cite{vaswani2017attention, ahmad2020transformerbased} & 44.6 & - \\
        
        \midrule
        FunCom1 & IR-Baseline  & 18.1 & BLEU-FC \\
        - & astattend-gru \cite{leclair2019neural} & 19.6 & - \\
        \midrule
        FunCom2 & IR-Baseline  & 18.2 & BLEU-FC \\
        - & astattend-gru \cite{leclair2019neural} & 18.7 \cite{alex2020improved} & - \\
        - & code2seq \cite{code2seq} & 18.8 \cite{alex2020improved}& - \\
        %- & BiLSTM+GNN-LSTM \cite{alex2020improved} & 19.05  & - \\
        - & code+gnn+BiLSTM\cite{alex2020improved}& 19.9 & - \\
        
        \midrule
        CodeNN & IR-Baseline & \emph{7.6} & BLEU-CN \\
        - & IR Iyer \textit{et al}. & 13.7 \cite{iyer2016summarizing} & - \\
        - & CodeNN \cite{iyer2016summarizing} & 20.4 & - \\
        
        \midrule
        DocString2 & IR-Baseline  & {\bf 32.7} & BLEU-ncs \\
        - & Transformer \cite{vaswani2017attention, ahmad2020transformerbased} & 32.5 \cite{ahmad2020transformerbased} & - \\
        \midrule
        DocString1 & IR-Baseline & {\bf 27.6} & BLEU-Moses \\
        - & Seq2Seq & 14.0 \cite{baroneDocstring}  & - \\
        %\cmidrule{2-4}
        %- & IR-Baseline & \textbf{39.2} & BLEU-1 \\
        %- & Hybrid-DeepCom & 15.6 \cite{wang2020trans3} & - \\
        %- & CoaCor & 25.6 \cite{wang2020trans3} & - \\
        %- & AutoSum \cite{DBLP:journals/corr/abs-1811-07234} & 25.3 & - \\
        %- & TranS$^{3}$ \cite{wang2020trans3} & 37.7 & - \\
        %\cmidrule{2-4}
        %- & IR-Baseline & \textbf{22.1} & METEOR \\
        %- & Hybrid-DeepCom & 6.1 \cite{wang2020trans3} & - \\
        %- & CoaCor & 9.5 \cite{wang2020trans3} & - \\
        %- & AutoSum \cite{DBLP:journals/corr/abs-1811-07234} & 9.3 & - \\
        %- & TranS\textasciicircum \cite{wang2020trans3} & 13.5 & - \\
        %\cmidrule{2-4}
        %- & IR-Baseline & 38.6 & ROUGE-L \\
        %- & Hybrid-DeepCom & 14.3 \cite{wang2020trans3} & - \\
        %- & CoaCor & 51.9 \cite{wang2020trans3} & - \\
        %- & AutoSum \cite{DBLP:journals/corr/abs-1811-07234} & 39.1 & - \\
        %- & TranS\textasciicircum3 \cite{wang2020trans3} & \textbf{51.3} & - \\
        %\cmidrule{2-4}
        %- & IR-Baseline & \textbf{166.6?} & CIDEr \\
        %- & Hybrid-DeepCom & 51.9 \cite{wang2020trans3} & - \\
        %- & CoaCor & 78.1 \cite{wang2020trans3} & - \\
        %- & AutoSum \cite{DBLP:journals/corr/abs-1811-07234} & 75.0 & - \\
        %- & TranS$^3$ \cite{wang2020trans3} & 87.2 & - \\
        %\midrule
        %MUSE & IR-Baseline  & aprox 36.9 & - \\
        \midrule
        NL de-en & IR-Baseline  & \emph{2.2} & SacreBLEU \\
        - & FAIR Transformer\cite{ng2019facebook} & 42.7\footnotemark{} & - \\
        \bottomrule
    \end{tabular}
    
    \caption{\label{tab:ir-results} Measurements of a simple information retrieval 
    baseline compared to various neural machine translation based methods. The scoring method we use
    mirrors the one used on the dataset (see Section \ref{sec:evalused}).}
    
\end{table}
\footnotetext{Value is for 2019 Model but on the 2018 test split}
\subsubsection{IR Results} 
\label{sec:IRresults}
We find that on most datasets the simple IR baseline approaches the neural models,
and exceeds it for DeepCom1, DocString1, and DocString2. 
However, IR does poorly on the WMT translation dataset, and also 
on CodeNN. In both cases, we speculate that this may reflect the relative level of redundancy
in these datasets. CodeNN is drawn from StackOverflow, which tends to have fewer duplicated questions; 
in the case of WMT, which is hand-curated, we expect there would be fewer duplications. 

Prior work~\cite{10.1145/3238147.3240471,Zhang2020RetrievalbasedNS} has used very sophisticated
IR methods. We cannot claim to supersede these contributions; but will point out that a very naive IR method
does quite well, in some cases better than very recently published
methods on datasets/dataset-variations which currently lack IR baselines. We therefore view IR baselines as important calibration on model performance; by trying such a simple baseline first one can help find pathologies in the model or dataset which require further exploration.

We also note that there is variation results. In DeepCom2f, which includes 10 cross-project folds, we observe a wide range results ranging from a BLEU-DC of {\bf 20.6} to {\bf 48.4}! This level
of variation across folds is a cause for concern...this suggests depending on the split, a model with higher capacity to memorize the training data might do better or worse, potentially muddling the results if only doing one split. Similarly we notice that between different versions of FunCom scores vary quite a bit; this variation may
confound measurement of actual differences due to technical improvements. 
\begin{minipage}{\columnwidth}
\begin{framed}
{\bf Recommendation: }{Since even naive IR methods provide competitive performance in many CCT datasets, they can be an important part for checking for issues in the new collection and new processing of CCT datasets. }
\end{framed}
\end{minipage}
% Notes
% The docstrings corpus seems to clip things to only include ones with
% between 2 and 400 tokens. This is how the 30k examples seem to disappear

\subsection{Calibrating BLEU Scores}

We now return our last research question, RQ~\ref{rq:calib}. How should we interpret the BLEU results reported
in prior work, and also the information retrieval BLEU numbers that we found (which are in the same range, see~\cref{tab:all-IR})?

\begin{figure}[h]
\includegraphics[width=0.6\columnwidth]{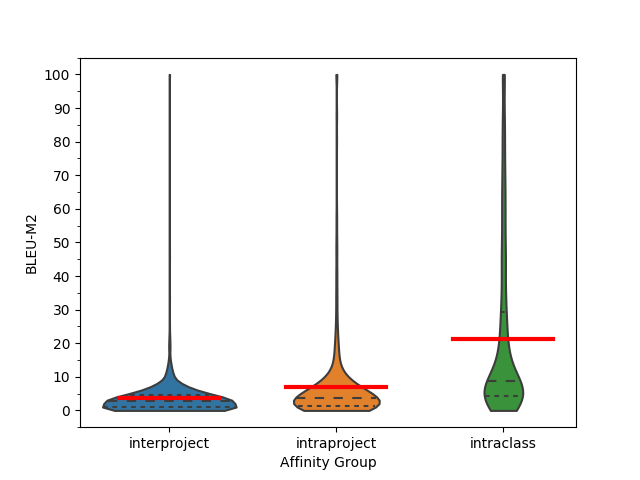}
\caption{The distribution of BLEU scores between affinity groups. Red lines represent the means (i.e. the Sentence BLEU), and the dashed lines represent
quartiles.}
\label{fig:violin}       
\end{figure}

To calibrate these reported BLEU scores, we conducted an observational study, using \emph{affinity groups} (AGs) of methods
that model different levels of expected similarity between the methods. For example, consider a random pair of methods,
so that both elements of the pair are \emph{methods from a different project}. This is our lowest-affinity group; we would expect
the comments to have very little in common, apart from both being utterances that describe code. The next higher
affinity group is a random pair of \emph{methods from the same project}. We would expect these to be a bit more
similar, since they are both concerned with the same application domain or function. The next higher level would
\emph{methods in the same class} which presumably are closer, although they would be describing different functions. 
By taking a large number random pairs from each of these affinity groups, and measuring the BLEU for
pairs in each group,  we can get an estimate of BLEU for each group. For our experiment, we picked the
1000 largest projects from Github, and then chose 5000 random pairs from each of
the affinity groups. For each pair, we randomly picked one as the ``reference" output, and the other as
the ``candidate'' output, and the BLEU-M2 score. We report the results in two different
ways, in ~\cref{fig:violin} and in ~\cref{tab:diffBLEU}. For intraclass, we do not take more than six random pairs from a single class. In all AGs, we 
removed all but one of the overloaded methods, and all getters and setters before our analysis. Without this filtering we see a difference of around 1-3  points.

First we describe~\cref{fig:violin} which shows the distribution of the BLEU scores in the 3 AGs. 
As might be expected, the inter-project AG shows a fairly low mean, around 3. The intra-project AG
is a few BLEU points higher. Most notably, the intraclass AG has a BLEU score around 22, which 
is close to the best-in-class values reported in prior work for some (but not all) datasets.

Note with the implemented experiment we cannot exactly compare these numbers, as each dataset is drawn from a different distribution.
Most CCT datasets provide the data in the traditional translation format of (source, target) pairs, making it difficult
to recover the other affinity group pairs of a given dataset example. This is why created our own new sampling based off of 1000 large Github repos. While not exactly comparable to existing datasets, new creations of CCT data could start with this simple 
affinity group baseline to calibrate the reported results.
 
Another, stronger, affinity grouping would be methods that are semantically equivalent. 
Rather than trying to identify such an affinity group ourselves by hand (which might
be subject to confirmation bias) we selected matched API calls from a recent
project, SimilarAPI~\footnote{\url{http://similarapi.appspot.com/}} by \citet{chensimilarapi} which used machine learning methods to match methods in different, but equivalent
APIs (e.g., \verb+Junit+ vs. \verb+testNG+). We extracted the descriptions of 40 matched pairs of high scoring matches from 
different APIs and computed their BLEU. 
We found that these BLEU scores
are on average about 8 points higher, with a mean around 32. 
This number should be taken with caution, however, since 
comments in this AG sample are substantially shorter than in the other groups.

\begin{framed}
{\bf Recommendation: }{Testing affinity groups can provide a baseline for calibrating BLEU results on a CCT dataset, as the simple trick of generating comments for a given method simply by retrieving the comments of a random other method in the same class possibly can approach SOTA techniques. }
\end{framed}

%\RSS{5}{Current SOTA BLEU scores, for the task of generating comments given code, are comparable to 
%the simple trick of generating comments for a given method by retrieving the comments of a random other
%method in the same class}

%%%%%% OverEdit
%\noindent{\bf Postscript:} (\emph{BLEU Variability})
%We noted earlier in~\Cref{sec:IRresults}, page~\pageref{sec:IRresults}, that there was
%considerable intrinsic variation within a dataset, simply \emph{across different folds}.
%This above affinity group experiment provided an opportunity to calibrate another BLEU variability, \emph{across different
%ways of calculating BLEU}. 
%%%%%%

\noindent{\bf Postscript:} (\emph{BLEU Variability})
We noted earlier in~\Cref{sec:IRresults}, page~\pageref{sec:IRresults}, that there was
considerable intrinsic variation within a dataset, simply \emph{across different folds}; we reported that measured
BLEU-DC in DeepCom2f ranged from 20.6 to 48.4; similar results
were noted in the different variants of FunCom. This above affinity
group experiment with IR provided an opportunity to calibrate
another BLEU variability, \emph{across different
ways of calculating BLEU}.

\begin{table}[h]
% Table autogenerated from affinity_data/analyze_affinity.py
\begin{tabular}{lr}
\toprule
 Function   &   intraclass \\
\midrule
 BLEU-FC    &        24.81 \\
 BLUE-CN    &        21.34 \\
 BLEU-DC    &        23.5  \\
 SacreBLEU  &        24.81 \\
 BLEU-Moses &        24.6  \\
 BLEU-ncs   &        21.49 \\
 BLEU-M2    &        21.22 \\
\bottomrule
\end{tabular}
\caption{The scores of samples of Java methods from the same class. }
\label{tab:diffBLEU}
\end{table}

%Second, these affinity group results allow us to characterize the variability that might arise
%from simply using different types of BLEU over these datasets. So for this reason,
We took the 5000-pair sample from the intraclass sample,
and measured the sentence BLEU for these pairs using the BLEU implementation variations used in the literature. The results are shown in~\cref{tab:diffBLEU}. The values range from around 21.2 to around 24.8; this range is actually rather high, compared to the gains reported
in recently published papers. This finding clarifies
the need to have a standardized measurement of performance. 
%1) How corpus BLEU and sentence BLEU for intra-class seems to be higher than most reported values in prior work. Prior work is split on what is actually reported. 

%2) Variance is very high here, but also in funcom. 

\begin{framed}
{\bf Observation:} Measurements show substantial variation. The version of BLEU chosen, and sometimes even the folds in
the training/test split, can cause substantial variation in the measured performance, that may confound
the ability to claim clear advances over prior work. 
\end{framed}

%% TODO: WHy is performance worse for CodeNN (David)
%% TODO: what is the BLEU confidence interval on a subsample of WMT for dataset of size roughly 2m when using IR.
%% TODO: what is the prevalence of duplicates what are the effects of duplications (where a code-comment pair occurs both in training and test) (David)
%% TODO: how is perormance of deep learningmodels affected by dups between training & validation. We need to do this only for DEEpcom & Leclaire, since codenn has no duplicates. First, do the easy thing, which is the effect of duplicates in test, and between training and test. Once this is done, then cap duplicates in trainign, and then evaluate on test. First part is more imporant. (Hari, David)
%% WRITING TODO: Please add file with text describing datasets, one para per dataset with cites (Hari)
%%  WRITING TODO: Add file with text explaining how plots and tables were produced (Hari)
%% WRITING TODO WRite intro, background & theory sections (Prem)
%% WRITING TODO Write text explaining how IR worked (David)
%% 
\begin{comment}

outline: 

1) How are Code-Comm datasets different from translation datsets?   Univariate & bivariate  
2) how is performance affected by frequently occurrring n-grams
3) How well does IR work
4) How is performance affected by duplicates, for IR & DL 
\end{comment}

\section{Discussion}
We summarize our main findings and their implications. 

\mypara{Comment Repetitiveness}
Our findings presented in Figure 1 show that comments in CCT datasets are far
more repetitive than the English found in the WMT dataset; figure 2 suggests
that this not merely a matter of greater vocabulary in distribution in comments,
but rather a function of how words are combined in comments. The highly-prevalent
patterns in comment have a substantially greater impact on the measured BLEU performance of 
models trained with this CCT data, as shown in Figure 3. A closer look at the
CCT datasets shows that trigrams such as \litcode{creates a new}, \litcode{returns true if}, 
\litcode{constructor delegates to}, \litcode{factory method for} are very frequent. Getting
these right (or wrong) has a huge influence on performance. 

\impara{Implications:} These findings suggest that getting just a few common patterns of comments right might
deceptively affect measured performance. So the actual performance of
comment generation might deviate a lot from measured values, much more so relative to natural language translation. 
Repetition in comments might also mean that fill-in-the-blanks approaches~\cite{sridhara2011automatically}
might be revisited, with a more data-driven approach; classify code first, to find the right
template, and then fill-in-the-blanks, perhaps using an attention- or copy-mechanism. 

\mypara{Input/Output Dependence}
When translating from one language to another, one would expect that more similar inputs
produce more similar outputs, and that this dependence is relatively smooth and monotonic. Our
findings in figure 4 and table 1, indicate that this property is indeed very strongly
true for general natural language outputs, but not as much for the comments. 

\impara{Implications:} Deep-learning models are universal high-dimensional continuous function approximators. 
Functions exhibiting a smooth input-output dependency, could be reasonably expected to be easier
to model. BLEU is a measure of lexical (token sequence) similarit); the rather non-functional nature
of the dependency suggested by figure 4 and table 1 indicate that token-sequence models that
work well for Natural language translation may be less performant for code; it may be that
other, non-sequential
models of code, such as tree-based or graph-based, are worth exploring further \cite{alex2020improved, 9031440}

\mypara{Baselining with IR}
Our experience suggests that simple IR approach provides BLEU performance
that is  comparable to current state of the art.

\impara{Implications} Our findings suggest that   a simple, standard, basic
IR approach would be a useful baseline for approaches to the CCT task. Especially
considering the range of different BLEU and tokenization approaches, this would be a
useful strawman baseline. 

\mypara{Interpreting BLEU Scores}
BLEU, METEOR, ROGUE etc are measures that have been developed for different task
in natural language processing, such as translation \& summarization, 
often after extensive, carefully designed, human subject studies. 
%\todo{different measures, different tasks}
Since BLEU is most commonly used in code-comment translation, we took an observational 
approach calibrate the BLEU score. Our results, reported
in~\cref{fig:violin} and~\cref{tab:diffBLEU} indicate that the reported BLEU scores are not
that high. 

\impara{Implications:} The best reported BLEU scores for the German-English translation tasks are currently are
in the low 40's. Our affinity group calibration suggests that on some datasets, the performance of models are comparable on average
to retrieving the comment of a random method from the same class. While this conclusion can't be explicitly drawn for a specific dataset without using the exact examples and processing from that specific dataset, but comparing results at an affinity group level can provide insight into minimum expected numbers for a new CCT datset.

\mypara{Learning From NLP Datasets}
We find that the current landscape of CCT datasets to be rather messy. There are often several different versions of the same dataset with different preprocessing, splits, and evaluation functions which all seem equivalent in name, but unless extra care is taken, might not be comparable.

However, some tasks in NLP do not seem to observe such variance within a task. We postulate this could be due to several reasons. For one, with the popularity of large open source repositories, it has become cheap and easy for a software engineering researcher to collect a large number of pairs of code and comments. This does not require hiring a human to label properties of text, and thus less effort might be taken on quality control compared to NLP data collection. Because researchers are domain experts in the datasets, they might be also more willing to apply their own version of preprocessing.

In addition, there are a wider array of tools to enforce consistency on various NLP tasks. For example the WMT conference on translation, a competition is ran with held out data and human evaluation. Other tasks, such as SQuAD\cite{DBLP:journals/corr/abs-1806-03822} for reading comprehension and GLUE\cite{DBLP:journals/corr/abs-1804-07461} for multitask evaluation allow for uploading code to a server which runs the proposed model on held out data. This ensure consistency in evaluation metrics and data.

We view adapting some these techniques as an interesting avenue for future work.

%% summarizing bullets. 
%% Future recommendations. 

\section{Threats to Validity}
%\todo{What measures are we sure about?}
%\todo{What are we not?}
Our paper is a retrospective, and doesn't propose any new tools, metrics,\emph{etc}, still
some potential threats exist to our findings. 

\impara{Fold Variance} With the exception of DeepCom2f we did not run measures over multiple folds or samples of the data. This makes it possible that there is variance in some of our reported numbers. 

\impara{The Affinity Benchmarks} When collecting affinity groups, we collect full methods and process them using a set of filters. This means that when comparing these numbers, they might not be dirrectly comparable to a specific dataset. The numbers are presented only as estimate of similarity of the affinity groups.

\impara{Replication Threat} Whenever we had to , we did our best to replicate, and measure the quantities we reported using
the same code as the previous work. Still, it is possible that we failed to comprehend some subtleties in the provided code, and this
may be a threat to our findings. 

\impara{Generalizability} We covered all the commonly used datasets and literature we could find. However, it may be that we have missed some where
cases our findings don't hold. 

\section{Conclusion}
In this paper, we described a retrospective analysis of several research efforts
which used machine learning approaches, originally designed for the task of natural language
translation, for the task of generating comments from code. We examined the datasets, 
the evaluation metrics, and the calibration thereof. 
Our analysis pointed
out some key differences between general natural language corpora and comments: comments
are a lot more repetitive. We also found that a widely used natural language translation dataset shows a stronger,
smoother input-output relationships than natural language.
Turning then to a popular evaluation metric (BLEU score) we found considerable
variation based on the way it's calculated; in some cases this variation exceeded
claimed improvements. Looking at calibration of the reported BLEU scores, 
first, we  found that simple 
off-the-shelf information retrieval offers performance comparable to that reported previously. 
Second, we found that the simple trick of retrieving a comment associated with a method in the same class
as a given method achieves an average performance comparable to current state-of-the-art. Our work suggests that future work in the area would benefit from a) other kinds
of translation models besides sequence-to-sequence encoder-decorder models b) more standardized
measurement of performance and c) baselining against Information Retrieval, and against some 
very coarse foils (like retrieving a comment from a random other method in the same class). 

Funding for this research was provided by National Science Foundation, under grant NSF 1414172, 
\emph{SHF: Large: Collaborative Research: Exploiting the Naturalness of Software}.

Source code and data will be made available
at {\color{blue}{\url{https://bit.ly/3lBDegY}}}
%online\footnote{\url{https://drive.google.com/drive/folders/1yceN-raSj3M12ajHc1aGdDAU1kRcC5jP?usp=sharing}}.

\bibliographystyle{ACM-Reference-Format}
%%% -*-BibTeX-*-
%%% Do NOT edit. File created by BibTeX with style
%%% ACM-Reference-Format-Journals [18-Jan-2012].

\end{document}